\documentclass[10pt,letterpaper,onecolumn,oneside]{article}
\usepackage[top=1in,left=1.25in,footskip=0.75in,marginparwidth=2in]{geometry}

\usepackage{booktabs} 
\usepackage{nopageno}
\usepackage{rotating}

\usepackage{upgreek}
\usepackage{xcolor}

\usepackage[utf8]{inputenc}

\usepackage[right]{lineno}

\usepackage{changepage}

\usepackage[aboveskip=10pt,labelfont=bf,labelsep=period,singlelinecheck=off]{caption}

\makeatletter
\renewcommand{\@biblabel}[1]{\quad#1.}
\makeatother

\usepackage{graphicx}
\usepackage{amssymb}
\usepackage{amsmath}
\usepackage{epstopdf}
\usepackage{comment}
\usepackage{enumerate}
\usepackage{color}
\usepackage{subeqnarray}
\usepackage{ulem}

\usepackage{lastpage,fancyhdr,graphicx}
\usepackage{epstopdf}
\usepackage{color}

\definecolor{Gray}{gray}{.25}

\usepackage{comment}
\usepackage{siunitx}
\usepackage{units}

\usepackage{epstopdf, epsfig}
\usepackage[english]{babel}
\usepackage{siunitx}
\usepackage{comment}
\usepackage{amsmath} 

\usepackage{graphics} 
\usepackage{epsfig} 

\begin{document}

\thispagestyle{plain}

\vspace*{0.35in}

\begin{flushleft}
{\Large
\textbf\newline{A simple criterion and experiments for onset of flocculation in kaolin clay suspensions}
}
\newline
\\
N. Rommelfanger$^1$, B. Vowinckel$^{1,3}$, Z. Wang$^2$, R. Dohrmann$^4$, E. Meiburg$^1$ \& P. Luzzatto-Fegiz$^1$

\bigskip
$^1${Department of Mechanical Engineering, University of California,
Santa Barbara, CA 93106}
$^2${Department of Physics, University of California, Santa Barbara, CA 93106}
$^3${Leichtwei\ss-Institut f\"{u}r Wasserbau, Technische Universit\"{a}t Braunschweig, 38106 Braunschweig, Germany}
$^4${Bundesanstalt f\"ur Geowissenschaften und Rohstoffe (BGR), Geozentrum Hannover, Stilleweg~2, 30655 Hannover, Germany}

\end{flushleft}

\begin{abstract}{Cohesive effects between fine-grained sediment particles greatly influence their effective settling rate and erodibility. Many studies have observed a qualitative difference in settling dynamics between clays in freshwater, where particles remain dispersed, and in saltwater, where aggregates form and settle rapidly. The critical coagulation concentration (CCC) of salt that separates the two regimes however remains under-investigated, even though knowledge of the CCC is crucial to understanding aggregation in settings such as estuaries, where large salt concentration gradients occur. Furthermore, no simple criterion exists to predict the CCC for clay suspensions. 
In this study, systematic experiments are performed to determine the CCC, by measuring transmitted light intensity through clay suspensions. To investigate the effect of ion valence, sodium chloride (NaCl) and calcium chloride (CaCl$_2$) are used. For kaolin clay, the results show a CCC of 0.6\,mM NaCl ($\approx 0.04$\,ppt NaCl $=$ 0.04 PSU), and of 0.04\,mM CaCl$_2$ ($\approx 0.004$\,ppt CaCl$_2$). Because these salinities are lower than those commonly observed in nature, these findings indicate that kaolin clay should flocculate in nearly all natural aquatic environments. Furthermore, due to the fact that tap water often has salinities higher than this threshold, these results imply that great care is needed in experiments, especially in large facilities where using distilled water is not feasible. 
In addition, a simple criterion to estimate the CCC for a kaolin clay suspension is derived. This criterion predicts that flocculation occurs at extremely low salt concentrations and is approximately independent of clay concentration, in agreement with the experimental observations and consistent with experimental evidence from the literature. }
\end{abstract}

\section{Introduction}
\label{intro}

Cohesive sediments comprising particles smaller than $\sim63\,\upmu \text{m}$ \cite{Mehta2013,Winterwerp2004} are ubiquitous in aquatic environments such as rivers \cite{Seminara2010-kq}, lakes and estuaries 
\cite{De_Swart2009-ro}, fisheries and coastal ecosystems, and benthic habitats. Cohesive sediments can experience flocculation, which greatly accelerates their settling rate and qualitatively alters settling dynamics. Closely related to this, cohesive forces also significantly alter the erodibility of the sediment grains. Owing to their plasticity and cohesive strength, clay minerals of grain size smaller than $\sim~2\,\upmu \text{m}$ are a key constituent of cohesive sediments \cite{Mehta2013, Shrestha2005,Winterwerp2004}. The cohesive interparticle forces responsible for flocculation are highly dependent on salt concentration, making it important to develop clay flocculation criteria that are both physically grounded, as well as tractable enough to be readily used in practical applications.

\begin{figure}
  \centerline{\includegraphics[width=.6\columnwidth]{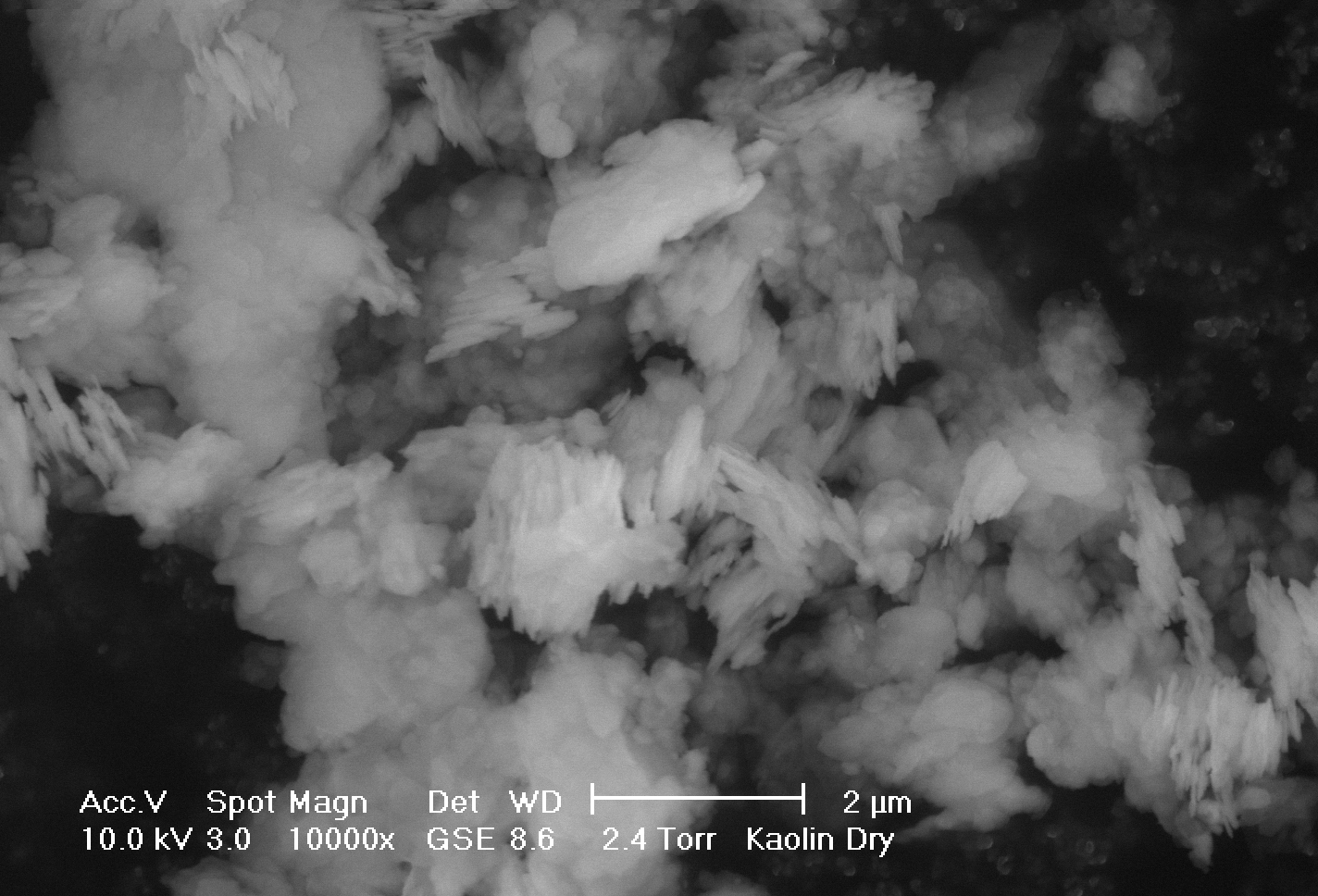}}
  \caption{Sample of dry kaolin powder used in the present experiments, imaged with an environmental scanning electron microscope.}
\label{fig:dry kaolin}
\end{figure}

The stability of clay suspensions (and therefore the onset of flocculation) is determined by the interplay between repulsive electrostatic forces and attractive van der Waals forces \cite{Urzay2010}. Kaolin clay is composed mainly of the mineral kaolinite together with other sedimentary minerals, such as quartz, feldspar, iron oxides and additional minerals in trace concentrations. Kaolinite is one of the most common clay minerals found in the environment and is widely used in laboratory models \cite{Nourmohammadi2011-ds,Yaghoubi_Sina2017-st,Yu_Wei-Sheng2000-gt}, and is therefore the focus of this study. Kaolinite particles are plate-like in shape (as shown by the micrograph in Fig.~\ref{fig:dry kaolin}), with negatively charged faces. An electric double layer forms around each particle when submerged in water, which leads to electrostatic repulsion between the like-charged particles. By adding positive ions in the form of dissolved salts, one may compress the double layer, decreasing the repulsive forces relative to van der Waals attraction and allowing aggregation to occur. The aggregates, or flocs, formed by this process settle faster than individual particles, sweeping up other flocs in their path, such that the suspended clay concentration decreases rapidly after the onset of flocculation \cite{Sutherland2015,Vowinckel2019}. The salt concentration at which the repulsive forces become smaller than the attractive forces is known in the literature on colloids as the `critical coagulation concentration' (CCC) \cite{vanOlphen1977}.

Although several investigations in environmental science have focused on clay suspensions, conflicting observations have been reported regarding the salinity required to trigger flocculation, with some researchers finding a significant influence of salinity on flocculation, and others measuring no influence. 
\cite{Tsai1987} measured the effect of shear on the flocculation of fine sediment (composed partially of kaolinite) from the Detroit River. They observed similar mean floc diameters for suspensions in both tap water and in well water with added salts. The authors concluded that small changes in freshwater composition do not strongly affect flocculation. 
\cite{Gibbs1989} observed increasing floc size in the freshwater Garonne and Dordogne Rivers (0.01 PSU) while moving downstream towards the Gironde Estuary.
\cite{Droppo1994} reported no correlation between major ion concentration and floc size in six rivers in southeastern Canada. 
\cite{Thill2001} measured the flocculation properties of suspended sediment taken from the Rhone River, and observed only a slight increase in mean floc size when the river water's salinity exceeded 25\,ppt NaCl. They concluded that salt-induced aggregation may not enhance flocculation at the river mouth. 
\cite{Huang2017} measured flocculation in the San Francisco Bay-Delta Estuary and observed similar floc sizes in the freshwater Sacramento River (0.1 PSU) as in saltwater.

Although field measurements are in disagreement about the effect of salt concentration, laboratory experiments have consistently reported that switching from freshwater to saltwater induces flocculation. 
\cite{Mietta2009} measured the effect of shear, pH, and salinity on the flocculation of laboratory-made or treated mud (treated with hydrogen peroxide) composed partially of kaolinite. They found that at a shear rate of 35\,s$^{-1}$ and pH~$=8$, mud with added salt formed larger flocs than mud without added salt. 
\cite{Sutherland2015} performed experiments with either freshwater or saltwater of at least 5\,ppt NaCl, and reported that kaolin clay suspended in saltwater settled more rapidly than kaolin clay suspended in freshwater. 
Although these laboratory studies show that flocculation increases upon adding salt, this increase in flocculation is not commonly observed {\it in situ} upon transitioning from saltwater to freshwater (for example, in estuaries). 
Indeed, flocculation is often observed in freshwater environments \cite{Eisma1991a}.
\cite{Fox2004} measured flocculation in the Po River Delta, and noticed that flocs had already formed upstream in the freshwater Po River, where kaolinite composes part of the sediment \cite{Boldrin1988}.

In summary, while laboratory experiments have observed a consistent difference in settling between pure freshwater and saltwater, field measurements and experiments with natural samples have provided contrasting results, with numerous studies reporting flocculation already taking place in freshwater, such that a negligible difference in flocculation was observed after the introduction of additional salt. Furthermore, laboratory studies of flocculation do not routinely report the salinity threshold for aggregation.

An expression for the CCC can be derived through Derjaguin Landau Verwey Overbeek (DLVO) theory, which calculates the energy balance of electrostatic repulsion and van der Waals attraction between plate-like particles \cite{Derjaguin1941, Verwey1948}. This derivation is, however, mathematically complex, and the final expression involves parameters that are difficult to measure and often unfamiliar to those outside of chemical engineering, such as the system's Hamaker parameter, the closest approach distance between particles, and the surface potential of the clay particles. Furthermore, there does not seem to be agreement over the 
definition of these input parameters and over how to measure them; as a matter of fact, the values selected by different users often vary by an order of magnitude \cite{Missana2000}. Moreover, despite its complexity, DLVO theory commonly requires additional modifications to completely describe the relevant physics; an example is given by the elastohydrodynamic corrections introduced by \cite{Urzay2010}. 
While manageable theories based on dimensional analysis have been developed for particle size distributions in clay suspensions \cite{Hunt1982}, the authors are not aware of similarly tractable models for the CCC.
A simple approach that could predict the value of the CCC (or even its order of magnitude) using easily found properties of the clay and ambient fluid would be very valuable in both laboratory and field applications. This would be especially valuable, for example, for predicting the onset of flocculation in estuaries, where a gradient of salt concentration from $\approx 0-35$ PSU occurs as rivers mix with coastal oceans, as well as for industrial applications, where countless processes involve ionic solutions of varying salinity \cite{Kotylar1996, Davis2010, Dobias2005}.

In this work, experiments are performed to accurately measure the CCC of kaolin clay dispersed in NaCl and CaCl$_2$ solutions, with a focus on testing the influence of very low salinities. Using scaling arguments, a tractable criterion is derived to predict the stability of clay suspensions. This criterion uses input parameters that are well-characterized for many types of clay, and provides a simpler alternative to DLVO theory. 

\section{Materials and Methods}\label{sec:materials}

\subsection{Clay Properties}
Kaolin clay was purchased as a dry powder from Sigma-Aldrich (product \#18616, typical density of 2.65\,g/cm$^3$  \cite{PubChem_2020-gm}). 
The quality of this specific kaolin clay was confirmed by X-ray diffraction of powdered specimen as well as oriented slides following treatments with air dried and ethylene-glycol-solvated conditions. Kaolinite ($\text{Al}_2 \text{Si}_2 \text{O}_5(\text{OH})_4$) dominated the clay as expected. Traces of other phyllosilicates such as mica and expandable clay minerals (smectite) along with minor amounts of quartz, feldspar and possibly traces of tridymite (SiO$_2$ polymorph) were also present. 

 The cation exchange capacity (CEC) of this specific kaolin clay was $3\,\text{mEq}/100\,\text{g}$, measured using the Cu-trien method as described in \cite{Dohrmann2012}. This CEC is within the range of $3-15\,\text{mEq}/100\,\text{g}$ from the classic study of \cite{Carroll1959}, which provides comprehensive CEC values for a wide variety of clays. The fact that the CEC value is at the lower end of the typical CEC range can partly be explained by the presence of admixtures of expandable clay minerals (smectitic), which implies that the kaolinite itself has a very low CEC. A PANalytical Axios XRF spectrometer was also used to determine the chemical composition of the kaolin clay, shown in Table \ref{tab:XRF}. The relatively large K$_2$O content of the kaolin clay was explained by the presence of K-feldspar.

\begin{table}
\caption{Chemical composition of kaolin clay used in the present experiments$^{a}$}
{\small
\centering
\begin{tabular}{rccccccccccc}
\hline

Constituent  & $\text{SiO}_2$ & $\text{Al}_2\text{O}_3$ & $\text{Fe}_2\text{O}_3$ & $\text{CaO}$ & $\text{MgO}$ & $\text{K}_2\text{O}$ & $\text{Na}_2\text{O}$ & $\text{MnO}$ & $\text{TiO}_2$ & $\text{P}_2\text{O}_5$ & $\text{SO}_3$ \\[3pt]
\hline
\% mass & 55.7                 & 39.6                                & 1.0                                 & 0.1                & 0.3                & 2.7                              & 0.1                               & 0.0                & 0.1                  & 0.2                                & $<0.01$ 

\\\hline
\multicolumn{12}{l}{$^{a}$Determined by X-ray fluorescence following the procedure described by \cite{Dohrmann2009}.}
\end{tabular}
\label{tab:XRF}
}
\end{table}

The specific surface area was determined as 8.6\,m$^2$/g, by using N$_2$ and Brunauer-Emmett-Teller (BET) analysis. A characteristic image of the kaolin clay powder from an environmental scanning electron microscope (ESEM, FEI Company XL30) is shown in Fig.~\ref{fig:dry kaolin}. 
Kaolin clay particles are arranged as typical booklets consisting of regularly-shaped kaolinite plates, and as dispersed particles thereof. The plate-like particles have two long axes and one short axis, and are typically $2\,\upmu \text{m}$ or less in width, and between 10-120\,nm in thickness \cite{Brady1996,Wan2002}. The particle size distribution of the kaolin clay was measured using X-ray granulometry (Micromeritics SediGraph 5100), showing that 62.4\% of the clay consisted of particles smaller than $2\,\upmu \text{m}$, and 93.7\% consisted of particles smaller than $6.3\,\upmu \text{m}$. 

\subsection{Fluid Properties}
As this study aimed to trigger and analyze flocculation in a precisely controlled manner, deionized (DI) water was used as the ambient fluid. The DI water had an electrical resistivity of 18\,M$\Upomega$-cm; this high resistivity was chosen to minimize the presence of ions in the ambient fluid that could contribute to flocculation. The salts consisted of NaCl and $\text{CaCl}_2$ from Sigma-Aldrich (product numbers 746398 and 793639 respectively). As is well known, NaCl is the most common salt found in seawater \cite{Lyman1940}. The choice of $\text{CaCl}_2$ enabled the observation of the effects of a divalent cation on flocculation, while still using chloride as the anion to limit the number of independent variables. Below, salinity is reported in units of molarity (M), that is, moles of salt per liter of solution. Molarity may be converted to parts per thousand (ppt) by multiplying the molarity by the salt's molar mass: this is 58.44\,g/mol for NaCl, and 110.98\,g/mol for CaCl$_2$. For NaCl solutions, practical salinity units (PSU, commonly used in oceanography and limnology) are equivalent to\,ppt.

\subsection{Experimental Apparatus}
Various methods have been used to measure colloid sedimentation or empirically define the critical coagulation concentration of colloids, and methods employing changes in optical properties are common for clay suspensions \cite{Auzeraislt1990,Garcia-Garcia2007,Goldberg1987,Kihira1992, Lagaly2003a,Palomino2005,Reerink1954, vanOlphen1977,VanZanten1992}. Higher suspended clay concentrations result in greater attenuation of light passing through a cuvette.  
Although measuring changes in transmitted light intensity is an indirect way to measure the CCC, its simplicity allows one to rapidly test a wide parameter space. 
Furthermore, determining the CCC by analyzing macroscopic changes in suspended clay concentration, as opposed to microscopic changes in particle size, is also useful for practical applications where the amount of clay in suspension is important, such as mining tailings \cite{Kotylar1996}, water treatment \cite{Davis2010}, and paper and ceramic production \cite{Dobias2005}.

\begin{figure}[t]
  \centerline{\includegraphics[width=0.6\columnwidth]{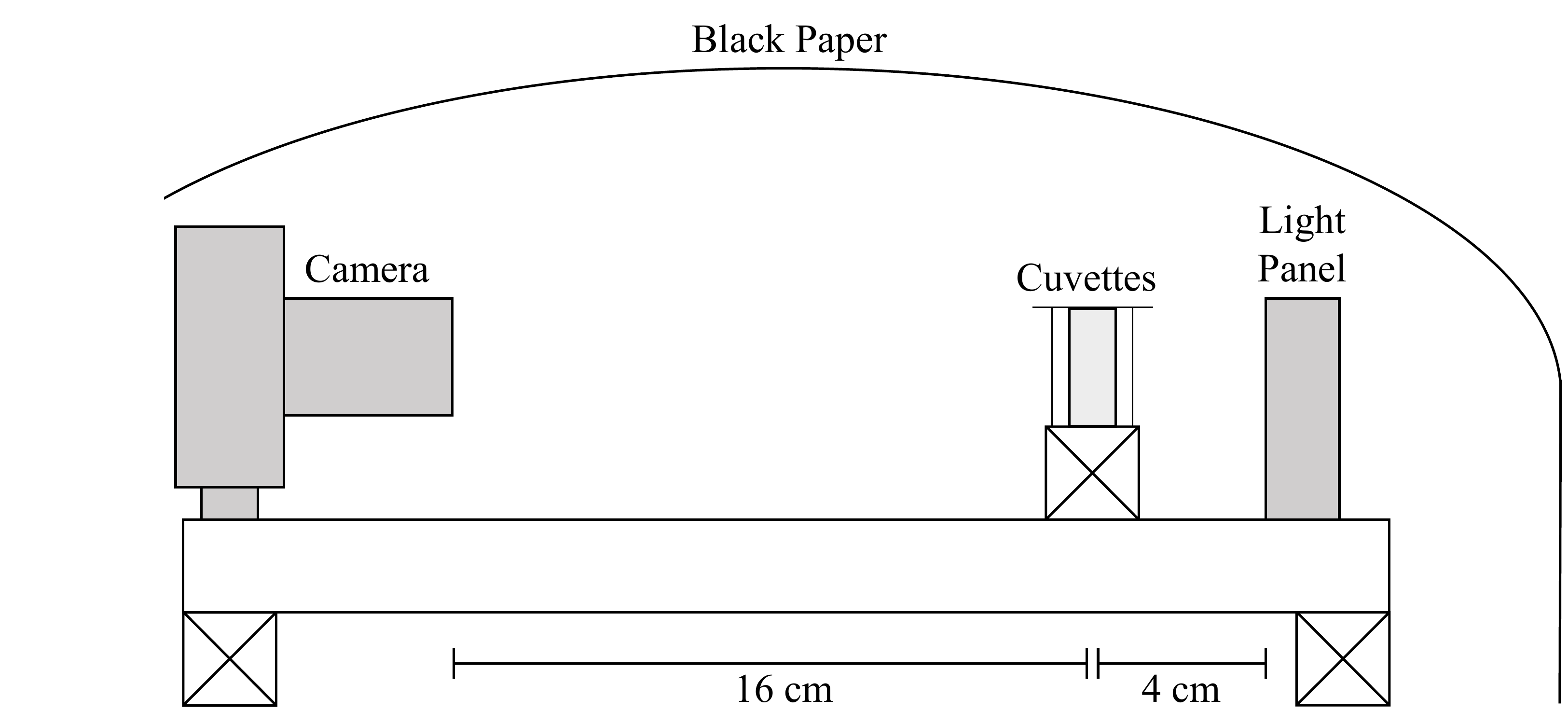}}
  \caption{The experimental setup. A camera (left) photographs a rack of five cuvettes (right-center), which are evenly illuminated from behind by a diffuse LED light panel (right). Black construction paper covering the entire setup prevents ambient light from reaching the camera's lens.}
\label{fig:setup diagram}
\end{figure}

The present investigation has been developed in preparation for microgravity experiments aboard the International Space Station (ISS). Due to operational constraints, the ISS setup uses the Binary Colloidal Alloy Test (BCAT) apparatus, which has been repeatedly employed to investigate the crystallization of colloid mixtures in microgravity \cite{Lu2007, Lu2009, Sabin2012} (ISS results involving cohesive sediment -- labelled `BCAT-CS' -- are presently being analysed). 
The laboratory setup employed for this study was therefore modeled after the BCAT apparatus, 
and is shown in Fig.~\ref{fig:setup diagram}. It comprises a camera (Nikon D2Xs with 60 mm lens, f/8, and 1/60 s exposure time) fixed on a rail 16 cm from a rack that holds five cuvettes. The cuvettes hold a volume of 10\,mm by 45\,mm by 4\,mm each, where the optical light path for illumination through the interior of the cuvette is 4\,mm, such that the field of view is 10\,mm by 45\,mm in width and height, respectively. Due to the small size of these cuvettes, one should not expect the settling velocities of initially dispersed clay suspensions (procedure described in detail below) to be independent of the container size \cite{michaels1962}. Preliminary tests were however conducted with different containers, which showed that the qualitative analysis on the triggering of flocculation was independent of container size (width, height) and container shape (cuboid, cylinder, cone). 
As described below, data was collected after the initial dynamics in the cuvette had ceased, in order to minimize the influence of reduced settling velocity. These small volumes enabled the rapid investigation of a wide parameter range of salinity and clay concentrations to determine critical thresholds for flocculation. Behind the cuvette rack, an LED light panel (Fotodiox C-218AS set to 14\% intensity and 4500 K temperature) uniformly illuminated the five cuvettes. The uniform lighting provided by the light panel eliminated the need for an additional diffuser. The entire setup was covered by black construction paper so that only the light passing through the clay suspensions reached the camera. 

\subsection{Experimental Procedure}
Results are reported for kaolin clay suspensions of three different concentrations: 4\,ppt, 20\,ppt, and 40\,ppt. For clay suspensions, ppt is defined as (clay mass) / (total mass)$\times 10^3$ (where both masses were, of course, measured using the same units). These clay suspensions allowed testing the effect of varying the clay concentration across one order of magnitude, while maintaining a sensible range of transmitted light intensity throughout the flocculation process. For each of these clay concentrations, several salt concentrations were tested, with additional measurements performed covering the concentration range around the onset of flocculation. 
For both NaCl and CaCl$_2$, these solutions ranged from $1.71\times10^{-3}$\,mM to $1.71\times10^{1}$\,mM, with a total of 8-12 different salt concentrations for each clay concentration. Because the focus of this investigation was on 
salt concentrations around the CCC, solutions with salinity higher than 17.1\,mM were not tested (note that 17.1\,mM NaCl = 1 PSU).

For each combination of clay concentration and salinity, a sample volume of 100\,mL was prepared. To remove contaminants, all clay samples were rinsed before testing by mixing the clay with DI water, allowing the clay to settle for at least 60 hours, and then removing the supernatant. This washing techniques is consistent with those used by other studies of clay suspensions \cite{Chen1998}. Furthermore, this washing procedure ensured that the clay had soaked in fluid for multiple days before experiments were conducted, thereby more closely matching the state of clay found in natural aquatic settings. Nevertheless, it was found that washed clay flocculated at approximately the same CCC as unwashed clay. 

To test whether small particles lost during the rinsing process had an impact on flocculation, an alternative washing procedure was also employed, whereby selected samples were centrifuged at 8000\,rpm for three minutes before removing the supernatant; this ensured that even the smallest particles were retained. The flocculation behavior of the centrifuged samples was the same as for samples that were not centrifuged.

Shortly before testing, all suspensions were shaken vigorously by using a mixing  
accessory attached to a 
reciprocating saw running at approximately 1500\,rpm for 15\,s. The saw had a stroke length of 1/2\,inch, thus accelerating the suspensions at approximately 130\,m/s$^2$. 

Once the sample was mixed, a pipette was immediately used to transfer small volumes of mixture into the glass cuvettes. Well-used glass cuvettes were used, as new glassware is often slightly alkaline \cite{Sigma-Aldrich}, which may affect the flocculation process at low salt concentrations. The cuvettes were re-agitated by hand once more just before placing them in the rack and beginning recording images with the camera. The loading procedure took about 40 seconds. The camera took a grayscale photograph every 20 seconds for ten minutes.

\subsection{Image Analysis}

\begin{figure}
  \centerline{\includegraphics[width=0.7\columnwidth]{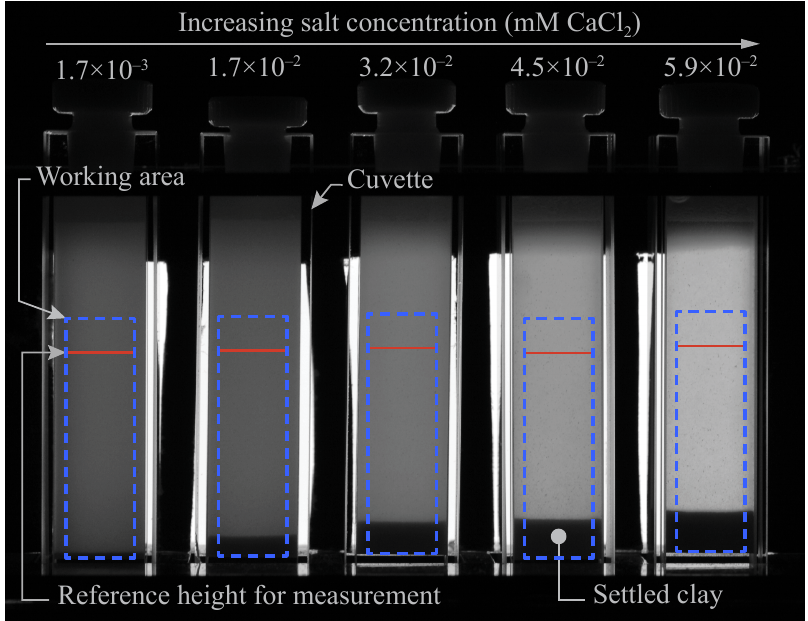}}
  \caption{Example photograph of settling progress at ten minutes for cuvettes with 20\,ppt kaolin clay suspensions. Salt concentration (CaCl$_2$) in the solutions increases from left to right. The areas marked by blue dashed lines are determined to be the ``working areas" in each cuvette. Light intensity was horizontally-averaged across the solid red lines, at 23.2\,mm above the bottom of the cuvettes. This averaged light intensity was used as a measure of suspended clay concentration.}
\label{fig:crops and line}
\end{figure}

In order to minimize the influence of solid boundaries on lighting and on the recorded settling behavior, a `working area' was delineated in the center part of the cuvette, as shown by the blue dashed lines in Fig.~\ref{fig:crops and line}.

As a metric of suspended clay concentration, the light intensity was horizontally-averaged at a height of 23.2\,mm above the bottom of the cuvette, for images acquired ten minutes after the start of the experiment (this height is marked by the red lines in Fig.~\ref{fig:crops and line}). This height was selected because, once a steady state was reached (as discussed below), this vertical location was distinct from both the settled clay on the cuvette floor and from the unevenly-lit region near the top of the cuvette. 

\begin{figure}
  \centerline{\includegraphics[width=0.7\columnwidth]{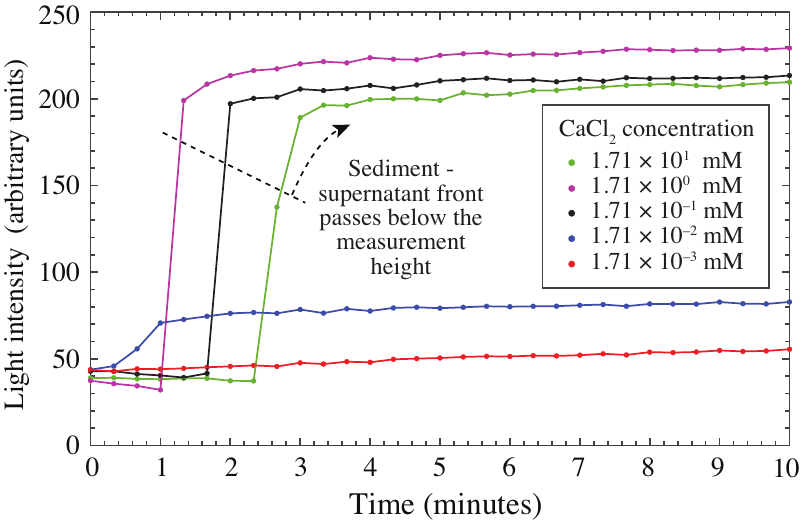}}
  \caption{Example of horizontally-averaged transmitted light intensity 23.2\,mm above the bottom of the cuvette vs. time for 40\,ppt kaolin clay suspensions with various CaCl$_2$ concentrations. Lines are included to guide the eye. The formation of a sediment-supernatant interface in the solutions with CaCl$_2$ concentrations at or above $1.71\times10^{-1}$\,mM, leads to a sudden increase in light intensity when the interface descends below the measurement height, between one and three minutes.
  }
\label{fig:intensity vs time}
\end{figure}

Figure~\ref{fig:intensity vs time} shows a time series of the light intensity at the reference height, for 40\,ppt kaolin clay suspensions with various concentrations of CaCl$_2$, ranging from $1.71\times10^{-3}$\,mM to $1.71\times10^{1}$\,mM. By ten minutes, the light intensities had settled into steady values. The formation of a defined sediment-supernatant interface in the solutions with concentrations of at least $1.71\times10^{-1}$\,mM CaCl$_2$ drastically affected the transmitted light intensity once the front descended below the measurement height. The formation of this front at high salt concentrations is consistent with observations in \cite{Kaya2006} and \cite{Sutherland2015}, among others.

These fronts descended below the reference height within the first three minutes; the light intensity remained relatively constant in all cuvettes afterwards. Thus, the suspensions were in a well-established steady-state regime at ten minutes, whereby one could observe how much clay had been left in suspension after the early sedimentation behavior.

The transmitted light intensity depended heavily on the clay concentration of the suspension under test. Therefore, in order to compare results with different concentrations on the same axes,
a normalized light intensity was calculated for each clay concentration, ranging from zero to one. 
To achieve this, the minimum light intensity was subtracted from the values at the end of the ten minute period (typically from the cuvette with the lowest salt concentration) for all cuvettes under test. The light intensities were then divided by the updated maximum light intensity at the end of the ten minute period (typically from the cuvette with the highest salt concentration). 

Motivated by a visual inspection of the results, the data was fit at the end of the ten minute period in MATLAB with an error function of the form: 
\begin{equation}
    I_{\text{norm}} = A\,\text{erf}\left(B\,\text{log}\left(\frac{s}{s_{\text{CCC}}}\right)\right)+D
\end{equation}
where $I_{\text{norm}}$ is the normalized light intensity, s is the salt concentration, and $A$, $B$, $s_{\text{CCC}}$, and $D$ are fit parameters. With this fit, the CCC is defined as the concentration at the mid-line of the error function curve, which marks the maximum rate of change in flocculation with increasing NaCl or CaCl$_2$ concentration. The CCC is often defined arbitrarily \cite{Goldberg1987,vanOlphen1977}, as flocculation changes over a finite range of salt concentrations from complete dispersion to complete aggregation. This definition using a fit parameter is robust, and is therefore appropriate for these CCC measurements.

\section{Experimental results}\label{sec:results}

\begin{figure}
  \centerline{\includegraphics[width=0.7\columnwidth]{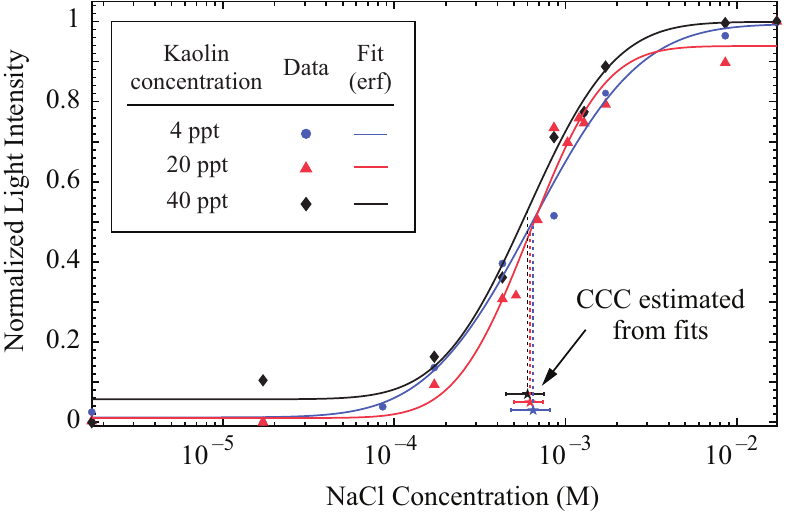}}
  \caption{Normalized light intensity versus NaCl concentration for 4\,ppt, 20\,ppt, and 40\,ppt kaolin clay solutions. The data is fit by an error function, and the CCC is defined as the concentration at the mid-line of the curve. The CCC of each curve is marked by a star of the same color below the curves, and 95\% confidence intervals for the CCC are marked by horizontal error bars. 
  }
\label{fig:nacl molarity}
\end{figure}

Figure~\ref{fig:nacl molarity} shows the results for NaCl solutions. We tested 4\,ppt, 20\,ppt, and 40\,ppt kaolin clay suspensions with NaCl concentrations ranging from $1.71\times10^{-3}$\,mM to $1.71\times10^1$\,mM. The CCC for each case is marked by a star below the curves (for clarity, the stars are placed at an arbitrary vertical coordinate); dashed lines are used to emphasize the mid-point of the transition in the curve. 
Horizontal error bars around the stars denote 95\% confidence intervals for the CCC as calculated by the curve fit. 
The CCCs for these kaolin clay suspensions lie around 0.6\,mM NaCl ($\approx 0.04$\,ppt NaCl $=$ 0.04 PSU).

Although the CCCs of the kaolin clay suspensions with NaCl appear to be somewhat lower than typically reported in the clay literature, this discrepancy is consistent with differences in 
methods of defining the CCC; for example, \cite{Wang2015} based their CCC estimates on the value of concentration beyond which no significant changes occur in the flocculation dynamics. By this definition, \cite{Wang2015} find a CCC of 3\,mM NaCl, which is consistent with the data in Fig.~\ref{fig:nacl molarity}, as all curves indeed level off around 3\,mM.

\begin{figure}
  \centerline{\includegraphics[width=0.7\columnwidth]{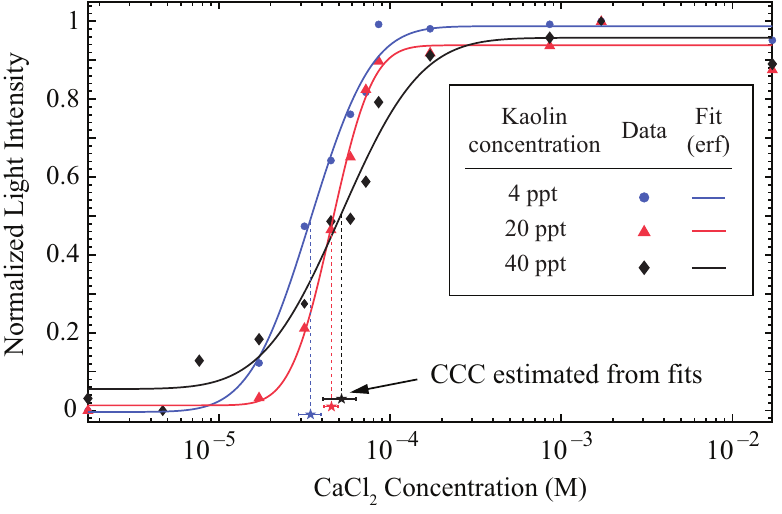}}
  \caption{Normalized light intensity versus CaCl$_2$ concentration for 4\,ppt, 20\,ppt, and 40\,ppt kaolin clay suspensions. The CCC of each curve is marked by a star of the same color below the curves, and 95\% confidence intervals for the CCC are marked by horizontal error bars. The CCCs for these solutions are over an order of magnitude smaller than the CCCs for NaCl solutions, ranging from 0.034\,mM for 4\,ppt kaolin clay to 0.052\,mM for 40\,ppt kaolin clay. 
  }
\label{fig:cacl2 molarity}
\end{figure}

Fig.~\ref{fig:cacl2 molarity} shows results for tests with kaolin clay and CaCl$_2$ solutions. Once more, tests were performed for 4\,ppt, 20\,ppt, and 40\,ppt  kaolin clay suspensions  with CaCl$_2$ concentrations ranging from $1.71\times10^{-3}$\,mM to $1.71\times10^1$\,mM. Here, the CCCs were found to range from 0.034\,mM CaCl$_2$ ($\approx 0.0038$\,ppt CaCl$_2$) for 4\,ppt kaolin clay to 0.052\,mM CaCl$_2$ ($\approx 0.0058$\,ppt CaCl$_2$) for 40\,ppt kaolin clay. Although the CCC in NaCl solutions did not appear to depend on clay concentration, a slight increase of the CCC with increasing clay concentration may be visible in the results for CaCl$_2$ solutions.

Incidentally, the fact that the 67 distinct cuvette time-series used to compile Figs.~\ref{fig:nacl molarity},~\ref{fig:cacl2 molarity} yield data that consistently collapses to the same curves suggests a good degree of reproducibility for these experiments.

\section{Theory}\label{sec:theory}
The large difference in CCC between NaCl and CaCl$_2$ solutions demonstrates the challenge of reliably predicting the onset of flocculation for a given clay suspension. This section presents a simple derivation of a dimensionless quantity that may be used to determine the effect of various clay and solution parameters on flocculation. For this purpose, the number of negative charges on suspended clay particles are compared to the number of positive charges surrounding the particles, and these two quantities are expressed in terms of parameters commonly known for a given clay. 

To begin, the clay particles are approximately represented as plates with major dimension $d$ and thickness $\tau$. For kaolinite particles, $d \approx 10 \tau$ \cite{michaels1962}. Thus, the edge surface area of a particle is nearly one order of magnitude smaller than the face surface area; for this reason, the edge surface area is neglected in the following analysis. 

The surface charge on the edge of the particles is also neglected, which is in accordance with early findings of \cite{Weiss1959}, who observed that exchangeable cations were mainly adsorbed on the hydroxyl-free basal planes (tetrahedral sheet) and to a lower amount on the hydroxyl-rich basal planes (octahedral sheet) rather than at the edges. This was confirmed by \cite{Kumar2017} using atomic force microscopy. Not only is the edge surface area much smaller than the face surface area, but the edge charge is pH dependent, and the point of zero charge (PZC) of kaolinite particles is near the pH measured for the solutions, at pH $=5.8$ \cite{Kretzschmar1998}. Thus, the edge charge is expected to be insignificant in this order-of-magnitude derivation.  

One wishes to obtain an expression relating the number of charges per surface area of the clay to the CEC. Because the CEC has units of number of charges per mass, it must be multiplied by the mass per area ratio of the particles to obtain the number of charges per surface area. 

\begin{figure}
  \centerline{\includegraphics[width=0.65\linewidth]{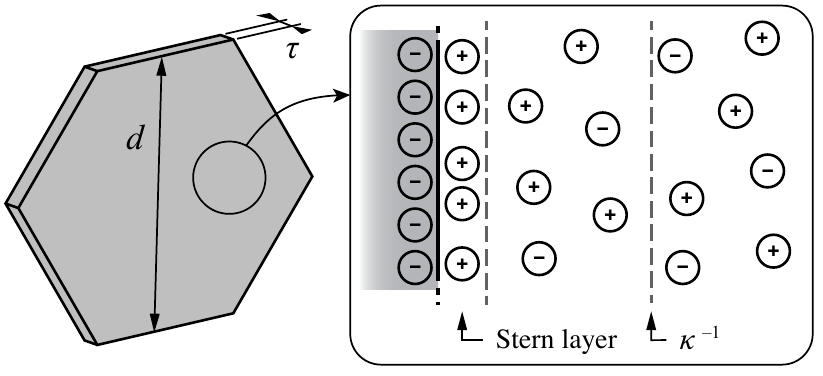}}
  \caption{Definitions of clay particle dimensions and of double layer thickness used in the model. The Stern layer consists of the cations adhered to the negatively-charged clay surface. The diffuse double layer comprises the cloud of ions beyond the Stern layer, where the Debye length $\kappa^{-1}$ is the characteristic thickness of the double layer, calculated with Eq.~(\ref{kappa}).}
\label{fig:debye}
\end{figure}

The area of one of the hexagonal faces of a clay particle is $A_\text{face} = \frac{\sqrt{3}}{2} d^2$, as illustrated in figure~\ref{fig:debye}. Because here the focus is on establishing a dimensionless group that characterizes the CCC, order-one constants are omitted, such that $A_\text{face} \sim  d^2$, where the symbol `$\sim$' denotes `order of magnitude'. Therefore the mass $m$ and surface area $A$ of a clay particle are given by:
\begin{equation}
    m = \rho_\text{clay} A_\text{face} \, \tau \sim \rho_\text{clay}\,d^2\,\tau,
\end{equation}
\begin{equation}
    A = 2\,A_\text{face} + \frac{6}{\sqrt{3}} d \tau \sim 2d^2,
\end{equation}
 where $\tau\ll d$ was used. Thus, one obtains:
 \begin{equation}\label{clay charge}
     \left(\frac{\text{\# charges}}{\text{surface}}\right)_{\text{clay}} \sim \text{CEC} \; \frac{m}{A}  = \text{CEC} \; \frac{\rho_{\text{clay}}\,\tau}{2} .
 \end{equation}
 
One must now solve for the number of ions in solution, per area of the particle they are surrounding. This requires defining a thickness for the `cloud' of ions enveloping the clay particle. A diffuse double-layer model is used, as illustrated in Fig.~\ref{fig:debye}, where the characteristic thickness of the counterion cloud surrounding the particle is the Debye length $\kappa^{-1}$ \cite{Mitchell1993}. The Debye length is calculated as: 
\begin{equation}\label{kappa}
    \kappa^{-1} = \sqrt{\frac{\epsilon_0 \epsilon_r k_B T}{2000 N_a e^2 I}}
\end{equation}
where $\epsilon_0$ is the permittivity of free space, $\epsilon_r$ is the medium's dielectric constant, $k_B$ is the Boltzmann constant, $T$ is the temperature, $N_a$ is Avogadro's number, $e$ is the charge of an electron, and $I$ is the ionic strength of the salt solution (in mol/L), defined below. While this expression is derived for the case of ions with equal valence, it is also a good approximation for the case of two ions with different valence \cite{Mitchell1993}. Additionally, while slight differences in the CCC of clays have been reported for different salts with the same valence number, these differences are generally less than one order of magnitude and thus fall within the error of this approximation \cite{rossington1999colloidal, li2021method, katz2013influence}.

The ionic strength $I$ of the salt solution expresses the number of charges per unit volume, and is given by:
 \begin{equation}
     I = \frac{1}{2} \sum_i z_i^2 s_i
 \end{equation}
where $s$ is the ion concentration (in mol/L) and $z$ is the ion's valence number, summed over each of the $i$ cations and anions in solution \cite{Garcia-Garcia2007}. To obtain the number of cations per area around the particle, one must multiply the ionic strength of the cations by the `cloud' volume per area ratio, or simply the thickness of the cloud. It is assumed that the cations relevant to screening the negatively-charged particle faces are found within a distance of order $\kappa^{-1}$ of the particle surfaces, and that the effect of anions is minimal within this region. Consistently with the earlier assumption about the edges of the clay particles, cations around the edges are also neglected, due to the small volume of the edge `cloud' relative to the face `cloud'. Thus, one considers a volume surrounding the faces of a clay particle, given by: 
 \begin{equation}
     V \sim 2 d^2 \kappa^{-1} .
 \end{equation}
 %
One thus obtains: 
 \begin{align}\label{salt charge}
     \left(\frac{\text{\# charges}}{\text{surface}}\right)_{\text{salt}} &\sim  z_{\text{cation}}^2 s_{\text{cation}}   \frac{V}{A}  \\ \nonumber &= \kappa^{-1} z_{\text{cation}}^2 s_{\text{cation}}.
 \end{align}
A dimensionless parameter $\Phi$ is now introduced, expressing the ratio of the effective number of cations in solution (\ref{salt charge}) to the number of anions on the clay surface (\ref{clay charge}): 
 \begin{align}
     \Phi &= \left. \left(\frac{\text{\# charges}}{\text{surface}}\right)_{\text{salt}}\, \right/ \, \left(\frac{\text{\# charges}}{\text{surface}}\right)_{\text{clay}} \\[5pt] \nonumber &=  \frac{2\kappa^{-1} z_{\text{cation}}^2 s_{\text{cation}}}{\text{CEC}\;\rho_{\text{clay}}\tau} .
 \end{align}
For NaCl, $z_{\text{cation}} =1$ and $s_{\text{cation}}  = s_{\text{NaCl}}$, giving:
 \begin{equation}
     \Phi = \frac{2\kappa^{-1}s_{\text{NaCl}}}{\text{CEC}\,\rho_{\text{clay}}\tau} .
 \end{equation}
For CaCl$_2$, $z_{\text{cation}} = 2$ and $s_{\text{cation}}  = s_{\text{CaCl}_2}$, giving:
 \begin{equation}
     \Phi = \frac{8\kappa^{-1}s_{\text{CaCl}_2}}{\text{CEC}\,\rho_{\text{clay}}\tau} .
 \end{equation}
When the dimensionless quantity $\Phi \sim 1$, the magnitude of the positive charge within $\kappa^{-1}$ of the faces of suspended clay particles is equivalent to the magnitude of the negative surface charge on the clay particles. Thus, it is expected that the surface charge will be effectively screened, and the repulsive force between clay particles will be approximately zero. This should lead to unimpeded flocculation.
It must be noted, however, that as salt concentration increases from zero, flocculation initially occurs when the attractive van der Waals force becomes greater than the repulsive electrostatic force (incidentally, this is equivalent to removing the energy barrier in DLVO theory \cite{vanOlphen1977}). This is expected to occur for a critical value of $\Phi$ below 1, as when $\Phi = 1$ the repulsive force is already effectively zero. This behavior would be consistent with experience from other areas of hydrodynamics, where transitions between regimes are observed when the ratios between stabilizing and destabilizing effects -- as expressed by relevant dimensionless groups -- are several orders of magnitude away from unity \cite{Avila2011,Chandrasekhar1961}. 

Below, a procedure is described for empirically determining the critical value $\Phi_{\text{CCC}}$ for the onset of flocculation. Once $\Phi_{\text{CCC}}$ is known, it is expected that a similar $\Phi_{\text{CCC}}$ should induce flocculation for other salts (or clays in the kaolin group). This will be verified with the experimental results. For many problems in geoscience, the simple dimensional analysis presented here can sidestep the need to employ significantly more complicated DLVO calculations to predict the necessary conditions for flocculation.

\begin{figure}[t!]
  \centerline{\includegraphics[width=\linewidth]{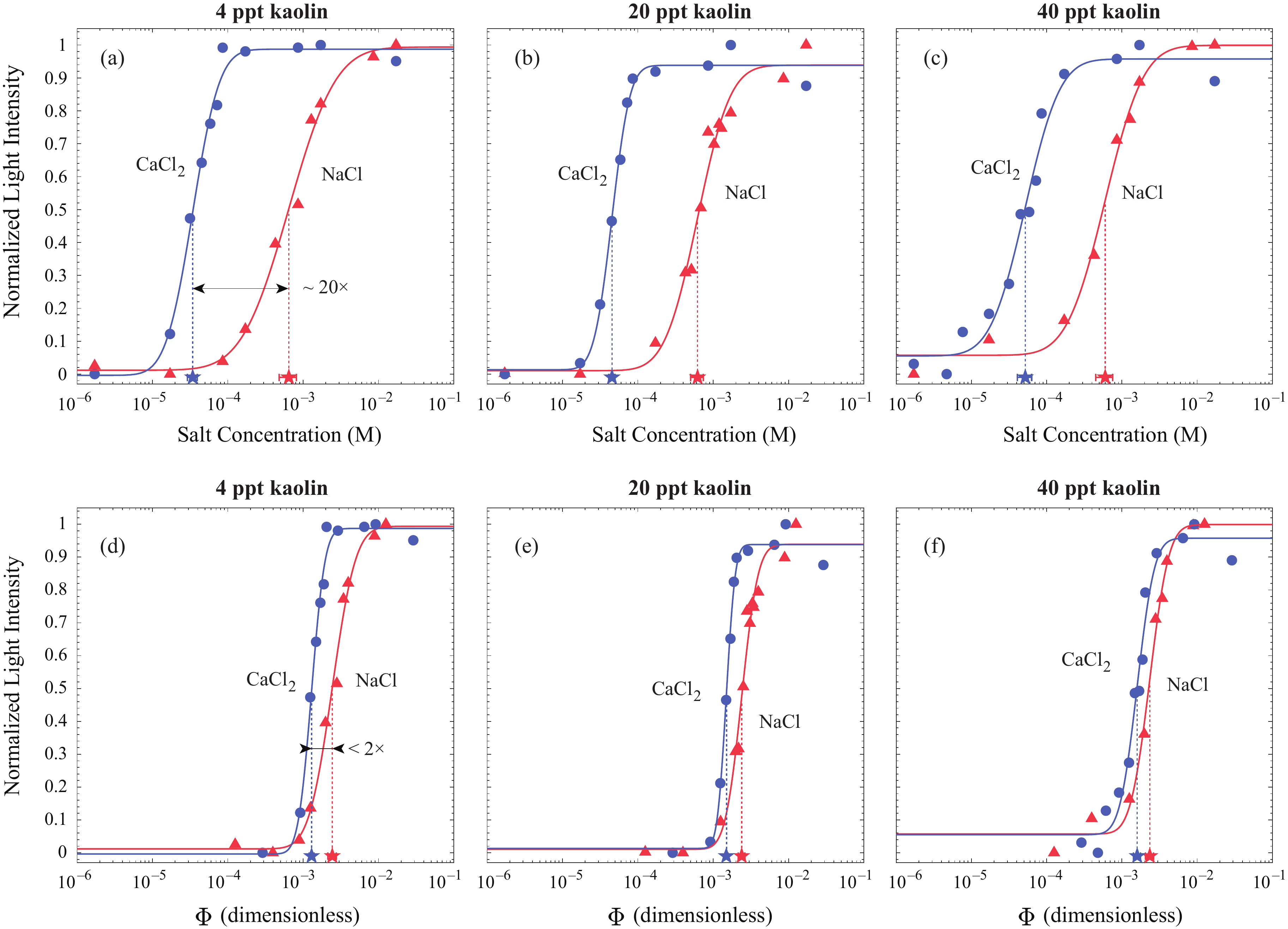}}
  \caption{Comparing plots of normalized light intensity vs. salt concentration (Figs.~\ref{fig:subplot comparison}a-c) and vs. the dimensionless group $\Phi$ (Figs.~\ref{fig:subplot comparison}d-f) for 4\,ppt, 20\,ppt, and 40\,ppt kaolin clay. The data is the same as was presented in Figs.~\ref{fig:nacl molarity} and \ref{fig:cacl2 molarity}. While the curves for NaCl and CaCl$_2$ are separated by over an order of magnitude when plotted against salt concentration, the curves are brought within a factor of 1.5-2 when plotted against $\Phi$. Thus, $\Phi$ can be used as an order-of-magnitude predictor of the onset of flocculation for various clay concentrations and salts of different valence. Here, $\Phi_{\text{CCC}}$ falls between $1.3 \times 10^{-3}$ and $2.4 \times 10^{-3}$, so it is estimated that the onset of flocculation will occur in clay suspensions when $\Phi = O(10^{-3})$.
  }
\label{fig:subplot comparison}
\end{figure}

\section{Using the parameter $\Phi$ to predict the onset of flocculation}\label{sec:discussion}
Fig.~\ref{fig:subplot comparison} shows a comparison of normalized light intensity plotted against salt concentration (Fig.~\ref{fig:subplot comparison}a-c) and against the dimensionless group $\Phi$ (Fig.~\ref{fig:subplot comparison}d-f) for 4\,ppt, 20\,ppt, and 40\,ppt kaolin clay suspensions. The data is the same as was presented in Figs.~\ref{fig:nacl molarity} and~\ref{fig:cacl2 molarity}, but now results for NaCl and CaCl$_2$ are shown on the same plot. The NaCl and CaCl$_2$ curves for all kaolin concentrations are separated by a factor of 12-18 when plotted against salt concentration.When plotted against $\Phi$, however, the NaCl and CaCl$_2$ curves draw within a factor of 1.5-2. This demonstrates the ability of the dimensionless group $\Phi$ to approximately collapse the curves of suspensions made with salts of different valence.

The empirically determined $\Phi_{\text{CCC}}$ is expected to be comparable for NaCl and CaCl$_2$ solutions. As mentioned above, when $\Phi \sim 1$, the repulsive force between clay particles will be effectively zero; however, because flocculation only requires that van der Waals forces can overcome electrostatic repulsion, flocculation should be triggered at some $\Phi_{\text{CCC}} < 1$. For the data sets shown in Fig.~\ref{fig:subplot comparison}, $\Phi_{\text{CCC}}$ falls between $1.3 \times 10^{-3}$ and $2.4 \times 10^{-3}$. Thus, is is estimated empirically that for a given clay suspension, flocculation will be triggered when $\Phi = O(10^{-3})$. 
For $\Phi \gg O(10^{-3})$, 
flocculation will occur, and suspended clay will rapidly clear from the supernatant. For $\Phi \ll O(10^{-3})$, 
flocculation will not occur and the suspension will be stable. The dimensionless group $\Phi$ should therefore be useful to those outside of the colloid community who want to predict the basic behavior of clay suspensions. Users may determine whether their suspension will be stable or unstable by substituting the proper numerical values for the materials in their suspension, all of which are well-characterized for common clays, with no need to employ significantly more complicated DLVO calculations. 

\section{Discussion}

\subsection{Implications for laboratory and environmental measurements of flocculation}
The results show that for nearly all cases outside of the laboratory, kaolin clay suspensions will naturally have a salinity greater than the CCC. 

As a matter of fact, these extremely small quantities of salt necessary to induce flocculation can often already be found in tap water. For example, water for UC Santa Barbara is supplied by the Goleta Water District, whose water quality testing throughout 2018 showed an average of 2.4\,mM Na$^+$ and 2.0\,mM Ca$^{2+}$ present in the surface water supply \cite{Goleta2019}. These concentrations are sufficient to induce flocculation in the kaolin clay, as the authors verified with experiments with tap water in the laboratory. This may explain why \cite{Tsai1987} did not observe a difference in flocculation upon adding more salt to tap water, as the salinity of the tap water (likely also from the Goleta Water District 30 years earlier) was already above the sediment's CCC. 

We also expect that \cite{Gibbs1989} observed an increase in floc size upon moving down the Garonne and Dordogne Rivers towards the Gironde Estuary because the very low salinity of the rivers (0.01 PSU $=0.17$\,mM NaCl) is below the CCC we measured for the kaolin clay solutions. The lack of correlation between salinity and floc size measured by \cite{Droppo1994} and \cite{Thill2001} may be explained if the (unreported) salinities of their river samples were already above their sediment's CCC. 

These measurements may also explain the similar floc sizes observed by \cite{Huang2017} in the Sacramento River and the San Francisco Bay-Delta estuary, as the river's salinity of 0.1\,PSU ($=1.7$\,mM NaCl) is above the CCC for the kaolin clay solutions. Additionally, the low salinity reported in \cite{Eisma1991a} is above the CCC we measured, suggesting that salt would have induced flocculation whether or not organic matter was present. 

Lastly, the salt content of the Po River has been repeatedly measured to contain calcium concentrations above the CCC of kaolin clay \cite{Pettine1994}, so although \cite{Fox2004} concludes that the presence of both calcium and organic matter lead to flocculation in the Po, the calcium's presence alone was likely sufficient to trigger flocculation.

Although many have observed a qualitative difference in settling between freshwater and saltwater in the laboratory, the extremely small magnitude of the CCC is often overlooked and unaccounted for in experimental design. For example, two otherwise identical laboratory studies of turbidity currents may find qualitatively different results depending on the cleanliness of the clay, on the water source (including effects due to variations in local tapwater properties), as well as on the presence of salt residues after past experiments involving even small amounts of dissolved salts. 

\subsection{Effects of ion valence}
The CCC that we measured for each kaolin clay concentration was over an order of magnitude smaller in CaCl$_2$ solutions than NaCl solutions: 18.9 times smaller for 4\,ppt kaolin clay suspensions, 13.7 times smaller for 20\,ppt kaolin clay suspensions, and 11.5 times smaller for 40\,ppt kaolin clay suspensions. This behavior is expected based on the Schulze-Hardy rule and subsequent developments, where the ratio of CCC$_{\text{CaCl}_2}$ / CCC$_{\text{NaCl}}$ is expected to range between 1/4 for low potentials and 1/64 for high potentials \cite{Overbeek1980}. 

These results also align well with other experiments, where it is common to observe at least one order of magnitude smaller CCC with CaCl$_2$ solutions than with NaCl solutions \cite{Hsi1962,Lagaly2003a, Swartzen-Allen1976}. 

The CCCs for the NaCl solutions in Fig.~\ref{fig:nacl molarity} do not appear to depend on clay concentration. This is consistent with previous results that have found that the CCC is independent of clay concentration when salts are not adsorbed in the Stern layer (e.g.~\cite{Lagaly2003a}; the Stern layer is illustrated in Fig.~\ref{fig:debye}). Monovalent cations do not experience as strong of an attraction to the negatively-charged clay surface as do multivalent cations, such as Ca$^{2+}$ \cite{Chan1984}, and it appears that there is little Na$^{+}$ adsorption in the Stern layer in this case. 

The CCCs for the CaCl$_2$ solutions in Fig.~\ref{fig:cacl2 molarity} increase slightly with increasing kaolin concentration, which may be explained by adsorption of Ca$^{2+}$ ions in the Stern layer. This adsorption of Ca$^{2+}$ leads to greater surface charge neutralization than is observed for Na$^+$, resulting in a very low CCC \cite{DeRooy1980}. 
If the CaCl$_2$ concentration is fixed and the clay concentration is increased, the effect of this charge neutralization becomes less noticeable in the presence of more clay particles with fully-charged surfaces. Thus, the CCC may be observed to increase with clay concentration in CaCl$_2$ solutions.

\section{Conclusions}\label{sec:conclusions}
The onset of particle aggregation is an important threshold for determining the dynamics of environmental flows. Here a simple criterion was presented, to predict the stability of a clay suspension using the dimensionless group $\Phi$, along with experimental results to empirically determine $\Phi_{\text{CCC}} = O(10^{-3})$. This newly defined dimensionless quantity $\Phi$ is an accessible tool for understanding flocculation in clay, providing a tractable alternative to the more laborious calculations of DLVO theory. 

It is expected that the use of the parameter $\Phi$ will allow for simple control of clay suspensions in aquatic systems ranging from the laboratory to the field. The experimental results show that the salt concentration needed to induce flocculation in kaolin clay is lower than what is typically observed even in tap water, and thus kaolin clay suspensions will flocculate in most environmental fluid mechanics applications.

\section*{Acknowledgments}

This work was supported by NSF ENG CBET Fluid Dynamics award \#1638156. N.R.~was supported by a UCSB MRL RISE award. B.V.~gratefully acknowledges the Feodor-Lynen scholarship provided by the Alexander von Humboldt foundation (Germany) and the German Research Foundation (DFG) grant VO2413/2-1. E.M.~furthermore gratefully acknowledges support by the Army Research Office under grant W911NF-18-1-0379. 

The authors thank Sage Davis at the UCSB Micro-Environmental Imaging and Analysis Facility for his assistance in operating the ESEM, and Otger Campas and Renwei Chen for access and assistance with the centrifuge. The authors also thank Brandon Grunberg, Kevin Hill, Balaji Subramanian, Meital Carmi, Omar Curiel, Rochishnu Chowdhury, Daniel Newton and Rena Yang for contributions to the development of the experimental setup and procedures. 

\bibliographystyle{unsrt} 

\bibliography{no_URL_CCC_Paper}

\end{document}